\documentstyle[]{article}

\def\momentum{\vec{\bf P}}   
\def\phat{\hat{\bf p}}            
\def\basis#1{{\bf e}_{#1}}     
\def\basisu#1{{\bf e}^{#1}}     
\def\eigen#1{{#1}_{\vec {\bf {\small P}}}}   
\def\svec#1{\vec {\bf {\small {#1}}}}   
\def\tpo#1#2{\vec{\bf {#1}},#2}    
\def\vector#1{\vec{\bf {#1}}}        
\def\threeplus{$3 \oplus 1\ $}  
\def\endo#1{{\tt End} \  {\bf R}^{#1}}    
\def\pseudo{{\bf \varepsilon}}         
\def\dim#1{$#1{\tt D}$}           
\def\nnn#1{\|\vec{\bf {#1}} \|}   
\def\ddd#1{\partial_{#1}^2}       
\def\expf#1{\exp \left({#1}\right)}   

\begin{document}

\title{Multivector Solutions to the Hyper-\\Holomorphic Massive Dirac Equation}

\author {\large William M. Pezzaglia Jr.}
\date{\small Dec 15, 1993 (Ver 2.4b)\\Preprint: clf-alg/pezz9302}
\maketitle

\begin{center}
Department of Physics\\
Santa Clara University\\
Santa Clara, CA 95053\\U.S.A.\\
{\small Email: wpezzaglia@scuacc.scu.edu}
\end{center}

\begin{abstract}
\noindent Attention is given to the interface of mathematics and
physics, specifically noting that fundamental principles limit the
usefulness of otherwise perfectly good mathematical general
integral solutions.  A new set of multivector solutions to the
meta-monogenic (massive) Dirac equation is constructed which form
a Hilbert space.  A new integral solution is proposed which
involves application of a kernel to the right side of the
function, instead of to the left as usual.  This allows for
the introduction of a multivector generalization of the
Feynman Path Integral formulation, which shows that particular
``geometric groupings'' of solutions evolve in the manner to which
we ascribe the term ``quantum particle''.  Further, it is shown
that the role of usual $i$ is
subplanted by the unit time basis vector, applied on
the right side of the functions.

Summary of talk, to appear in: {\it Proceedings of the 17th
Annual Lecture Series in the Mathematical Sciences, April 8-10,
1993, University of Arkansas}, {\tt Clifford Algebras in Analysis},
John Ryan editor (CRC Press 1994).
\end{abstract}

\section{Introduction}
As a physicist, I am like `a stranger in a strange land' where familiar names
have different meanings.  It become increasingly clear that the practitioners
hypercomplex analysis, and those of multivector physics do not occupy the same
``Clifford'' space.  This is I believe a manifestation of the classical chasm
which unfortunately often exists between the disciplines of {\tt applied
mathematics} and {\tt theoretical physics}.  My attempts to bridge this gap
by conversation at this conference often ran agound.  Lacking background
(or interest) in physics, many mathematicians are reluctant to be drawn
into unfamiliar territory (and of course visa versa).  For brevity I will
continue from my own point of view, but it should be clear that
many of the statements I make are equally true if you interchange the words
``mathematician'' and ``physicist''.  Further,  any parochial statements are a
function of deliberate hyperbolism.

There is perhaps an unconscious assumption that the interface of mathematics
and physics is a {\it no-man's land}, which is not to be ventured into under
threat of being accused, by ones colleagues, of defection to the other side.
For example, the theoretical physicist that treads too close to the border
runs the risk of having his work dismissed as ``mere reformulation'', or worse
as {\it just mathematics}, i.e.\ not `real' physics.  The point is well taken.
While learning more math will enable the physicist to communicate better with
mathematicians, it will not necessarily lead to new or even better physics.
This is because physics is much more than the study of the subset of
mathematical equations that is isomorphic to physical phenomena.  On the other
hand, mathematicians are far more than nit-picking grammarians
that clean up the details after physicists have done the \underline{real}
work of creating new language (e.g.\ Newton's calculus) as a side effect
of their struggles to comprehend the universe.

Rather than dismiss practitioners of the opposition because they lack
sophistication in our own language, it behooves us to acknowledge that
mathematics and physics are \underline{different disciplines} with distinct
agendas and criteria.  An example in point was the `open problem book' of
the conference, in which mathematicians posed concise questions for
which an answer is yet unknown (or even known to exist, as in
Fermat's last theorem) but presumably can be solved through clever
logical deduction.  It was difficult to come up with an analogous
``physics'' question.  As recently discussed by Romer\cite{Romer}
such questions can usually only be answered by experimental
verification (e.g.\ does the {\tt Top Quark} exist), or an application
of a {\it physical principle}.
It was the downfall of Galileo that the `new science' had new physical
laws that could not be derived by logical extension or verified by
mathematical proof.  Given this, it is somewhat of mystery that
physical principles can be so conveniently expressed in mathematical
form\cite{Wigner}.

The goal of this paper is twofold.  Specifically we consider the more
general {\tt meta-monogenic equation} (in non-Euclidean spacetime),
which is of more interest (to
physicists) than the more limited monogenic equation, yet has not
received nearly as much attention in Clifford analysis.  Secondly,
this paper will be used as a vehicle to address the {\it interface} of
mathematics and physics, by providing associations between
mathematical concepts and physical principles.

\section{Algebraic Notation}
We seek to use a geometric algebra which will describe the empirically
known properties of physical spacetime.  Specifically, the mathematical
structure chosen should encode physical principles (e.g.\ Lorentz covariance).

\subsection{Classical Galilean Space and The Pauli Algebra}
Before the theory of relativity, physicists thought ``space'' was
intrinsically three dimensional.
Physical quantities [such as the electric field vector
$\vector{E}=\vector{E}(\vector{X},t)$\ ]
would be represented as a function of a position
{\it Gibbs vector} $\vector{X}$ and {\it scalar} time t,
$$(\vector{X},t)=x\sigma_1+y\sigma_2+z\sigma_3+t\sigma_0,
\eqno(2.1)$$
where $\sigma_0=1$ is the unit scalar.  The \threeplus notation of
eq.\ (2.1) matches our perception, which insists upon seeing time
as a scalar, different from \dim{3} space.  Equivalently, we say that the
domain $(\vector{X},t)$ belongs to ${\bf R}^3\oplus {\bf R}$; it is an
{\it ordered pair} of a three space {\it Gibbs vector} and a {\it scalar}.
This structure (i.e.\ setting $\sigma_0=1$) was assumed by many of the
contributors at this conference (n.b.\ plenary presentations).

The Clifford algebra generated by the three mutually anticommuting
basis elements $\{\sigma_1,\sigma_2,\sigma_3 \}$ is $\endo{3}$,
commonly called the {\it Pauli algebra}, with matrix
representation of $2 \times 2$ complex matrices ${\bf C}(2)$.
While this algebra and the \threeplus notation is
sufficient for many physical applications (see for example
Baylis\cite{Baylis}),
there are points where it will however lead to ambiguities
(n.b.\ time reversal), which can only be resolved with the full \dim{4}
concept.

\subsection{Minkowski Spacetime and Majorana Algebra}
The {\it principle of Lorentz invariance} demands that $(\vector{X},t)$
is an element of a \underline{four}  \underline{dimensional} domain,
rather than \threeplus.
Algebraically this means that the basis element associated with time
[(e.g.\ $\sigma_0$ of eq.\ (2.1)] must anticommute with the other
three mutually anticommuting basis vectors.  In order to avoid
confusion, we will introduce a new set of symbols: $\{\basis{\mu} \}$
as the basis vectors of \dim{4} space, with $\basis{4}$ (instead of
$\basis{0}$) as the fourth basis element.  Hence the coordinate
{\it four-vector} will be expressed in boldface lower case (no overhead arrow),
$${\bf x} = (x,y,z,t) = x^{\mu} \basis{\mu}=
x\basis{1}+y\basis{1}+z\basis{3}+t\basis{4}, \eqno(2.2)$$
where the Einstein summation notation is assumed on the repeated
index: $\mu =1,2,3,4$.

The {\it metric} of time is empirically known to be of the opposite
sign as that of the \dim{3} positional portion.  Mathematically, we would
say that the orthogonal space is either ${\bf R}^{3,1}$ or ${\bf R}^{1,3}$,
corresponding to metric signatures $(+++-)$ or $(---+)$ respectively
(physicists refer to these respectively as the {\it east coast} or
{\it west coast} metrics).  In the former case, the norm of the
position four-vector would be,
$$\|{\bf x}\|= x_{\mu}x^{\mu} = x^2+y^2+z^2-t^2, \eqno(2.3)$$
where $x_4=-x^4=t$, but $x_i=x^i$ for $i=1,2,3$. Physically, this
hyperbolic nature is interpreted as a manifestation of the {\it principle of
causality}.  For an ``event'' at the origin to be the source or {\it cause}
of another event at ${\bf x}$, the separation four-vector between them
must have norm $\|x\|^2 < 0$ in the ${\bf R}^{3,1}$, space (or $\|x\|^2 > 0$
in the ${\bf R}^{1,3}$ space).  Equivalently, we way that the vector
connecting an event with its cause is {\it timelike}, or lies inside the
{\it light cone} (the hypersurface for which $\|x\|^2=0$).

The Clifford algebra generated by the four basis vectors in the
``east coast'' $(+++-)$ metric is known as the {\it Majorana algebra},
with matrix representation ${\bf R}(4) = \endo{3,1}$.
This algebra is different to that generated by the other metric
choice (used for example by Hestenes\cite{Hestenes82}), which has
inequivalent matrix representation ${\bf H}(2) =\endo{1,3}$, i.e.\
$2 \times 2$ quaternionic matrices\cite{Porteous}.
Regardless, either algebra has 16 basis elements, but
\underline{neither} contains the global commuting $i$, which is
usually required by physicists for standard relativistic quantum
mechanics\cite{Drell64}.  Those theories use the ``classic''
{\it Dirac algebra}, which has \underline{five} anticommuting generators
$\{\gamma^{\mu}\}$, corresponding to complex matrix representation
${\bf C}(4)= {\bf C}\otimes {\bf H}(2) =  {\bf C}\otimes {\bf R}(4)$.

The ``projection'' or {\it spacetime split} from
${\tt 4D \rightarrow 3\oplus 1}$
can be encoded algebraically in ${\bf R}(4)$,
$$-{\bf x} \ \basis{4} = \vector{X} + t, \eqno(2.4)$$
where the \dim{3} Pauli algebra generators $\sigma_j$ of eq.\ (2.1)
are related to the Majorana generators: $\sigma_j = \basis{4}\basis{j}$
for $j=1,2,3$.

\subsection{Automorphisms and Conservation Laws}
Fundamental to physical theories are {\it conservation laws}, e.g.\
conservation of energy.  These are intimately related via Noether's
Theorem\cite{Drell65} to the physical principles being
invariant under certain transformation (e.g.\ rotations).  The ``allowed''
symmetries should correspond mathematically to those generated by the
automorphsim group of the geometric algebra.

Of particular interest are the orthogonal transformations associated
with hyperbolic rotations between space and time.  These have the
physical interpretation of connecting two reference systems, one at
rest and the other moving at velocity $\vector{V}={d\vector{X} \over dt}$.
A particle of mass $m$ at rest in the ``moving'' frame, will be perceived
as moving at velocity $\vector{V}$ in the ``rest'' frame, with
{\it four-vector momentum} {\bf p},
$${\bf p}=p^{\mu}\basis{\mu}=
{\cal L} \ m\basis{4}\ {\cal L}^{-1}, \eqno(2.5a)$$
$${\cal L}=\expf{ {r \hat{\beta}} \over 2} =
{ {-{\bf p}\basis{4} + m} \over \sqrt{2m(E+m)}}=
{\momentum + E + m \over \sqrt{2m(E+m)}},\eqno(2.5b)$$
$$E=p^4 = -\basis{4}\cdot{\bf p}, \eqno(2.5c)$$
$$\vector{P}=E\vector{V}=\basis{4}\wedge{\bf p}, \eqno(2.5d)$$
$$\tanh(r) = \nnn{V} = E^{-1} \nnn{P}, \eqno(2.5e)$$
$$\hat{\beta} = {\vector{V}\over \nnn{V}}=
{\vector{P}\over \nnn{P}}. \eqno(2.5f)$$
The term ${\cal L}$ is called the (half-angle) {\it Lorentz Boost operator},
where $r$ is the {\it rapidity} associated with the velocity.  The unit
bivector $\hat{\beta}$ points in the direction of the \dim{3} velocity
$\vector{V}$, or \dim{3} momentum $\vector{P}$.  The factor of
$\basis{4}$ in the middle of eq.\ (2.5b) causes the spacetime split
[see eq.\ (2.4)] of the four-momentum, where the fourth component
$p^4$ is physically interpreted as the {\it energy} $E$ in eq.\ (2.5c).

We define the following algebra involutions,
$${\cal T}(\Gamma)=-\basis{1}\basis{2}\basis{3}\ \Gamma
\ \basis{1}\basis{2}\basis{3}, \eqno(2.6a)$$
$${\cal P}(\Gamma)=-\basis{4}\ \Gamma\ \basis{4}, \eqno(2.6b)$$
which correspond geometrically to reflections.  The first inverts
timelike geometry, hence is called {\tt Time Reversal} in physics.
Basis elements which are invariant under this involution (e.g.\
$\basis{1}\basis{2}$) are called {\it spacelike}, while those which
acquire a minus sign (e.g.\ $\basis{1}\basis{4}$) {\it timelike}.
We will have more use for the second, the contrapositive inversion of
\dim{3} space called the {\tt Parity Transformation}\cite{Drell64},
under which many (but not all) physical laws are invariant.
The composition of the two is the {\it main involution} of the algebra,
$${\cal PT}(\Gamma)={\cal TP}(\Gamma)=-\basis{1}\basis{2}\basis{3}\basis{4}
\ \Gamma\ \basis{1}\basis{2}\basis{3}\basis{4}, \eqno(2.6c)$$
inverting the odd geometry (i.e.\ vectors and trivectors).

The anti-involutions associated with eq.\ (2.6b) and (2.6c) are
respectively the ``dagger'' and ``bar'' (equivalently the {\it Hermitian}
and {\it Dirac} conjugates respectively).  They are related,
$$\overline{\Gamma}=-\basis{4}\ \Gamma^{\dagger} \ \basis{4}, \eqno(2.7a)$$
$$\overline{\bf e}_{\mu}=-\basis{\mu}, \eqno(2.7b)$$
$$\basis{4}^{\dagger}=-\basis{4},\ \ \ \basis{j}^{\dagger}=+\basis{j}
\ (j=1,2,3), \eqno(2.7c)$$
where the ``bar'' reverses the order of all elements and inverts the
basis vectors.  The multivector bilinear form $\overline{\Psi}\Psi$ has the
advantage of being Lorentz invariant (as the Dirac bar conjugate
operation inverts the bivector generators of Lorentz transformations).
On the other hand, the scalar part of this form is not in
general positive definite, in contrast to the form $\Psi^{\dagger}\Psi$.
Both forms will in general have non-scalar portions.

\section{Functional Solutions of the Massive Dirac Eqn.}
The adjective ``massive'' should be superflourious, because historically
Dirac was describing the electron, a particle with mass.  It is
therefore a source of confusion for the physicist to encounter the
use of the term ``Dirac equation'' in Clifford analysis sometimes applied
to the {\it monogenic} equation $\Box\Psi=0$, rather than
eq.\ (3.2a) below.  When $\Psi$ is restricted to be
a bivector, the monogenic equation would be called the (sourceless)
{\tt Maxwell equation}, describing the spin-one \underline{massless}
{\sl photon} (i.e.\ electromagnetic waves).
If $\Psi$ is projected onto a minimal
ideal (i.e.\ a {\it column spinor}), it would describe a spin-half (again
massless) {\sl neutrino}, often called the {\tt Majorana equation}.  These
are special cases which have many interesting properties, unfortunately
which vanish once mass is included, or a source added (which makes
the equation non-homogeneous).  In this paper, we will address the
more general {\it meta-monogenic} or ``hyper-holomorphic'' situation,
in the \underline{non}-Euclidean spacetime metric case, which so far
has not received much attention in Clifford analysis.

\subsection{Relativistic Quantum Wave Equations}
The massive {\it meta-harmonic} equation over the spacetime
domain ${\bf R}^{3,1}$ is known as the
{\tt Klein-Gordon equation}\cite{Drell65},
$$(\Box^2 - m^2)\ \phi({\bf x})=0, \eqno(3.1)$$
where $\Box^2=\ddd{x}+\ddd{y}+\ddd{z}-\ddd{t}$
is the {\tt d'Alembertian} (the \dim{4} generalization of
the {\tt Laplacian}, but in non-Euclidean spacetime).  Note
that some authors use the symbol $\Box$ without the square
for the d'Alembertian, which makes the use of the same symbol
for the factored Dirac operator ambiguous.  Physically, eq.\ (3.1)
is interpreted to be the relativistic generalization of
Schr\"odinger's quantum wave equation.  Historically, $\phi$
was a complex scalar function describing a charged spinless
particle.  If however the function $\phi$ is a vector, then eq.\ (3.1)
is called the {\tt Proca equation}, describing a spin-one
particle (the ``vector boson''equivalently electromagnetism with mass).

Factoring the Klein-Gordon operator minimally requires four
anticommuting algebraic entities $\basis{\mu}$,
$$(\Box - m)\ \Psi({\bf x})=0, \eqno(3.2a)$$
$$\Psi({\bf x})=(\Box + m)\ \phi({\bf x})=
(\basisu{\mu}\partial_{\mu}+m)\ \phi({\bf x}), \eqno(3.2b)$$
$$\{\basis{\mu},\basis{\nu}\}=2g_{\mu \nu}, \eqno(3.2c)$$
where $\basis{\mu}=g_{\mu\nu}\basisu{\nu}$, and
$\Box=\basisu{\mu}\partial_{\mu}$ is the \dim{4}
spacetime gradient (we hesitate to call it the ``Dirac operator'',
because this term is sometimes used by physicists to refer to the
second quantized wavefunction $\Psi$).  Following
Greider\cite{Greider}, we note that the {\it metric tensor} $g_{\mu\nu}$
\underline{must} have the diagonal values
$g_{11}=g_{22}=g_{33}=-g_{44}=+1$ of the {\it east coast}
signature $(+++-)$ if the use of the commuting $i$ is excluded.
Note that this means that eq.\ (3.2a) and most of the following
results of this paper are inaccessible if the other metric
$(+++-)$ with inequivalent algebra
${\bf H}(2)$ is used, as assumed by Hestenes\cite{Hestenes82}.

\subsection{Meta-Monogenic Functions}
Real solutions (no $i$) to the {\it meta-harmonic} Klein-Gordon
eq.\ (3.1) are of the form:
$\eigen{\phi}({\bf x}) \sim \cos{(p_{\mu}x^{\mu})} $.
Substituting this eigenfunction into eq. (3.1) gives the
{\it characteristic equation},
$$ m^2 = -p_{\mu}p^{\mu} = E^2-\| \momentum \|^2 , \eqno(3.4a)$$
known in this case as the {\it Einstein relation} between mass $m$, energy
$E$ and vector momentum $\momentum$.  Note for a given mass and vector
momentum, the energy could be positive or negative, the latter is unphysical.
We shall define energy to be positive,
$$\pm p^4 = E = +\sqrt{m^2 + \| \momentum \|^2}. \eqno(3.4b)$$

Substitution $\phi=\cos{(p_{\mu}x^{\mu})}$ into eq.\ (3.2b) yields a
multivector solution to the Dirac eq.\ (3.2a) which can be
expressed in the exponential form\cite{Pezz8301}\cite{Ross},
$$ \eigen{\Psi}(x) =
\exp{(-\phat \ p_{\mu}x^{\mu})}\  \Lambda, \eqno(3.5a)$$
$$ \phat=m^{-1} {\bf p} =
m^{-1}\  p^{\mu}\basis{\mu}, \eqno(3.5b)$$
$$ {\phat}^2 = -1, \eqno(3.5c)$$
where $\Lambda$ is an arbitrary geometric factor (the ``spin'' degrees
of freedom).  The unit four-velocity $\phat$ plays the role of the usual
$i$ as the generator of quantum phase (when multiplied sinistrally,
i.e.\ on the left).  However, each momentum eigenfunction has its own
{\it unique} generator $\phat$.
Note that eq.\ (3.5a) is invariant under the replacement of
${\bf p} \rightarrow {\bf -p}$, hence we can restrict
eq.\ (3.4b) to $p^4=+E$ without any loss of generality,
hence: $p_{\mu}x^{\mu}=\vector{P}\cdot\vector{X}-Et$.

Consider a solution which is a linear combination of
eigenfunctions of eq. (3.5a), restricted to $\Lambda=1$.  This is a
reasonable construct to consider because of the {\it superposition
principle} of quantum theory: {\sl the sum of any two physically
interpretable solutions will be a reasonably interpretable solution}.
Although each eigenfunction separately will have a unit quadratic form,
$\overline{\Psi}\Psi=1$, the new solution
will display non-scalar  ``interference'' terms due to the differing
generators $\phat$.  These have been interpreted as a possible useful
description as a source for mesonic interactions\cite{Pezz9302}.

\subsection{Multivectorial Hilbert Space}
In order to have a probabilistic interpretation of quantum theory,
we need a positive definite norm.  A restricted choice of factor
$\Lambda$ in eq.\ (3.5a) will make the eigenfunctions unitary\cite{Pezz8301},
$$\Lambda=\Lambda_{\svec{P}}=
\sqrt{m \over{(2\pi)^3 E}}\  {\cal L}, \eqno(3.6)$$
where the term ${\cal L}$ is the Lorentz Boost operator of eq.\ (2.5b).
Using the property of the Lorentz operator,
${\bf p}{\cal L} = {\cal L} m \basis{4}$ from eq.\ (2.5a), the
unimodular meta-monogenic multivector eigenfunction can
be expressed in the alternate form,
$$\eigen{\Psi} (x) =
\Lambda_{\svec{P}}\ \exp{(-\basis{4}\ p_{\mu}x^{\mu})}, \eqno(3.7a)$$
$$=(\Box + m)\ \eigen{\Phi}(x), \eqno(3.7b)$$
$$\eigen{\Phi} (x)=
{\exp{(-\basis{4}\ p_{\mu}x^{\mu})}\over
(2\pi)^{3 \over 2} \sqrt{2E(E+m)}}. \eqno(3.7c)$$

Now we see that each eigenfunction has the {\it same} generator of
quantum phase, the (positive parity) unit time vector $\basis{4}$ when
multiplied {\it dextrally} (right-side applied).  Further the
meta-monogenic eigenfunction $\eigen{\Psi}({\bf x})$ can be written
in terms of a multivector solution $\eigen{\Phi} ({\bf x})$ to the
Klein-Gordon equation, again complex $\basis{4}$.  This differs
from the work of Hestenes\cite{Hestenes82} and Benn\cite{Benn}
which both interpret the negative parity geometric pseudoscalar as playing
the role of $i$.  Eigenfunctions in that form will not obey the
standard parity relation, which ours do,
$${\cal P}\left( \eigen{\Psi} (\tpo{X}{t}) \right) =
-\basis{4}\ \eigen{\Psi} (\tpo{X}{t})\ \basis{4} =
+ {{\Psi}_{-\svec{P}}} (\tpo{-X}{t}). \eqno(3.8)$$

The unitary meta-mongenic eigenfunctions are orthonormal and complete,
$$\int d^3{\vec{\bf X}}\ \Psi_{\svec{K}}^{\dagger}(\tpo{X}{t})
\ \eigen{\Psi}(\tpo{X}{t}) = \delta_{\svec{K}, \svec{P}}, \eqno(3.9a)$$
$$\delta(\vector{X}-\vector{Y}) = \int d^3\vector{P}\ \eigen{\Psi}(\tpo{X}{t})
\ \eigen{\Psi}^{\dagger}(\tpo{Y}{t}).  \eqno(3.9b)$$
Hence these functions are the restricted subset of the solution space
that forms a {\it \underline{Multivector} Hilbert space}, where again
$\basis{4}$ ({\it dextrally} applied, i.e.\ multiplied on the right
side) plays the role of the usual $i$.  A general solution in this
function subspace (e.g.\ a ``quantum wavepacket'') can be expanded,
$$\Psi (\tpo{X}{t}) =\int d^3\vector{P}
\ \eigen{\Psi} (\tpo{X}{t})\ {\cal C}_{\svec{P}}, \eqno(3.10a)$$
$$ {\cal C}_{\svec{P}} =
\int d^3\vector{X}\ \eigen{\Psi}^{\dagger}(\tpo{X}{t})
\  \Psi(\tpo{X}{t}), \eqno(3.10b)$$
where the coefficients $ {\cal C}_{\svec{P}}$ can be thought of
as ``3D scalars'', complex in $\basis{4}$ and \underbar{must} appear on
the right side of eq.\ (3.10a).  Note these coefficients turn out
to be independent of time $t$.

The full 16-degree-of-freedom solution to eq.\ (3.2) is of mixed
parity and can be written,
$$\Psi (\tpo{X}{t}) =\int d^3\vector{P}
\ \eigen{\Psi} (\tpo{X}{t})\ {\cal C}_{\svec{P}}
\ [f_{+}(\vector{P},\vector) +
\pseudo\ f_{-}(\vector{P})], \eqno(3.11a)$$
$$\pseudo=\basis{1}\basis{2}\basis{3}\basis{4}, \eqno(3.11b)$$
where $f_{\pm}$ are multivector functions with four degrees of
freedom on the basis set consisting of the scalar and the
timelike trivectors:
$\{1, \pseudo\basis{1}, \pseudo\basis{2}, \pseudo\basis{3}\}$.
These represent the `spin' degrees of freedom, of both parities
(indicated by the sign subscript).  This full solution can be
interpreted as an algebraic representation of isospin doublet of
Dirac bispinors\cite{Pezz9201}\cite{Pezz9302}.

\section{Integral Meta-Monogenic Solutions}
In general, the first-order monogenic equation: $\Box\Psi=0$\ [or
the generalized Dirac eq.\ (3.2a)] more directly lends itself to
integral solution than the associated second-order harmonic
equation [or the generalized Klein-Gordon eq.\ (3.1)].  For example,
time independent (i.e.\ \dim{3}) ``static'' electromagnetism
can be easily re-expressed in Cauchy integral form, even when
non-homogeneous source terms are included.  A typical electrostatic
example would be to calculate the electric field $\vector{E}$
at any point inside a region knowning the field on the boundry
(equivalently knowing the surface charge density $d\sigma=
\vector{E}\cdot d\vector{A}$) and the distribution of (scalar)
charge $\rho=\nabla \vector{E}$ inside the region\cite{Jackson}.
On the other hand, the Bergman kernel would only provide a method
of calculating the electric field at a point in the region, if
one already knows the complete solution everywhere inside the
region.  I am not aware of any classical problem which makes
use of this circular relationship, however it possibly could
be used to generate a perturbation series approximation.

There are additional constraints imposed by fundamental physical
laws which further limit the usefulness of integral formulations
in describing tangible phenomena.  For example, consider the \dim{3}
{\tt Helmholtz equation}: $(\nabla^2+\lambda^2)\phi(\vector{X})=0$,
of which integral solutions to the factored meta-monogenic form have
been recently treated by Shapiro\cite{Shapiro}.  This should
provide a good alternative derivation of Kirchhoff diffraction
theory, however it will still suffer from the practical
limitation of the function only being approximately known
on the boundry\cite{Jackson}.  Alternatively, this equation could be
physically interpreted as a factored form of the
time-independent quantum
{\tt Pauli equation} (see recent exposition by Adler\cite{Adler}).
However, one is then ontologically constrained by the quantum
principle which states that only the modulus $\rho=\phi^{\dagger}\phi$
of the wavefunction $\phi=\sqrt{\rho}\exp(i\theta)$ can be measured;
the absolute quantum phase $\theta$ cannot be known.  Hence it will
be impossible to perform a Cauchy boundry integral except for very
special situations.  This has influenced the way that physicists
approach or ``formulate'' quantum problems.

\subsection{The Propagating Kernel}
One common scenario is to \underline{assume} a physical system's initial
configuration, {\sl even though it cannot be directly measured},
[e.g. $\Psi(\vector{X},t^{\prime})$ known
for all $\vector{X}$ at $t^{\prime}=0$], and then to ask how the
system must change with time.  The {\it Time Evolution Operator}
$U(t;t^{\prime})$ is defined,
$$\Psi(\tpo{X}{t})=
U(t;t^{\prime})\ \Psi(\tpo{X}{t^{\prime}}), \eqno(4.1a)$$
$$U(t,t^{\prime})=U(t-t^{\prime})=
\exp{[-(t-t^{\prime})\ (\vector{\nabla} - \basis{4} m)]}, \eqno(4.2b)$$
where $\vector{\nabla}=\basis{4}\wedge\Box = \basis{4}\basisu{j}\partial_j$
is the \dim{3} gradient.  It will be assumed that $t>t^{\prime}$,
otherwise one would be talking about `retrodiction', a different
problem.  An integral representation of eq.\ (4.1a) is,
$$\Psi(\tpo{X}{t}) =
\int d^3\vector{X}^{\prime}\ F(\tpo{X}{t};\tpo{X^{\prime}}{t^{\prime}})
\ \Psi(\tpo{X^{\prime}}{t^{\prime}}),\eqno(4.3a)$$
$$F(\tpo{X}{t};\tpo{X^{\prime}}{t^{\prime}})=
U(t,t^{\prime})\ \delta(\vector{X}-\vector{X^{\prime}}),\eqno(4.3b)$$
where the {\it Dirac Delta Function} $\delta(\vector{X})$ is a
distribution of measure one at the origin and zero elsewhere.  The kernel
$F({\bf x,x}^{\prime})$ of eq.\ (4.3b) is known in physics as the
{\it transformation function}\cite{Abers}, or sometimes the
{\it Hyugen Propagator} since eq.\ (4.3a) is a description of
Hyugen's principle: {\tt each point on a wave front may be regarded
as a new source of waves}.  The integral is over the \dim{3}
hypersurface defined by $t^{\prime}=$constant, however it could be
generalized to any {\it spacelike} surface, i.e. a surface made of
points that are not {\it causally connected} to each other, and
further that $t$ of eq.\ (4.11a) is greater than the $t^{\prime}$
of each point on the hypersurface.  This amounts to
requiring that the surface cannot have any ``kinks'' in it that would
have the surface locally slip inside the ``light cone''.  The
mathematician would equivalently say it must be a {\it Lipschitz surface}.

By inspection of eq.\ (4.3b), the kernel is function only of
the \underline{difference} of the coordinates:
$F(\tpo{X}{t};\tpo{X^{\prime}}{t^{\prime}})=F({\bf x},{\bf x^{\prime}})=
F({\bf x}-{\bf x^{\prime}})$.  Substituting eq.\ (3.9a) for the
delta function in equation (4.3b) gives a closed integral form,
$$F(\tpo{X}{t};\tpo{X^{\prime}}{t^{\prime}})=
\int d^3\vector{P}\ \eigen{\Psi}(\tpo{X}{t})
\ \eigen{\Psi^\dagger}(\tpo{X^{\prime}}{t^{\prime}}), \eqno(4.4a)$$
$$=\int {d^3 \vector{P} \over {(2\pi)^3 E}}
\ \exp{[-\phat\ p^{\mu}(x_{\mu}-x^{\prime}_{\mu})]}
\ {{\bf p}\ \basis{4}}, \eqno(4.4b)$$
$$= -(\Box +m)\ \Delta ({\bf x - x^{\prime}})\ \basis{4}, \eqno(4.4c)$$
$$\Delta({\bf x})=
\int {d^3\vector{P}\over {(2\pi)^3 E}} \ \sin{(p^{\mu}x_{\mu})}. \eqno(4.4d)$$
{}From eq.\ (4.4a) it is clear that: $F^\dagger({\bf x,y}) = F({\bf y,x})$.
The function $\Delta({\bf x})$ is given by Bjorken\cite{Drell65}
to be related to a regular Bessel function inside the light cone.
Clearly it is a solution everywhere to the Klein Gordon eq.\ (3.1),
and so it follows from eq.\ (4.4c) that the propagator kernel is
{\it left meta-monogenic},
$$(\Box - m)\ F({\bf x,x^{\prime}})=0, \eqno(4.5a)$$
$$(\Box + m)\ F({\bf x^{\prime},x})=0, \eqno(4.5b)$$
where $\Box$ operates on ${\bf x}$, but not on ${\bf x^{\prime}}$.
What we have derived is a special case of the {\it Cauchy kernel}.
The usual closed boundry integral reduces to eq.\ (4.3a) when one
assumes that $\Psi(\vector{X},t)\rightarrow 0\ {\rm for:}
\ \nnn{X}\rightarrow \infty\ {\rm or}\ t\rightarrow \infty$.
In quantum theory, one sidesteps the unmeasurable phase by assuming
an `initial' solution at $t^{\prime}\rightarrow -\infty$ which is
a coherent plane wave.  This approximates for example the
collimated beam of electrons shooting down the linear accelerator
at Stanford.

\subsection{Green Function}
Another common situation is to know the function over a restricted
spacelike surface (e.g. a long thin cylinder, such as an antenna),
but different than our previous example, the time dependence of
the function is known (e.g. harmonic).  This will
require the use of a {\it Green function} which satisifies:
$(\Box - m)\ G({\bf x}) =\delta^4({\bf x})$.  When the use of $i$
has been excluded, the inhomogeneous part will take the form,
$$G({\bf x}) = \int{ {d^4{\bf k}} \over {(2\pi)^4 [m^2 + {\bf k}^2]}}
[m \cos(k^{\mu}x_{\mu}) + {\bf k} \sin(k^{\mu}x_{\mu})], \eqno(4.6a)$$
$$=(\Box + m)\ \Theta (t)\ \Delta ({\bf x}), \eqno(4.6b)$$
$$=\Theta (t)\ (\Box + m)\ \Delta ({\bf x}), \eqno(4.6c)$$
$$=\Theta (t)\ F({\bf x})\ \basis{4}, \eqno(4.6d)$$
$$=\Theta (t) \int {d^3\vector{P} \over {(2\pi)^3 E}}
\ {\bf p}\ \exp(-\phat \ p^{\mu} x_{\mu}), \eqno(4.6e)$$
where $\Theta(t)$ is a step function, which takes on the value of
$+\frac{1}{2}$ for $t>0$, and $-\frac{1}{2}$ for $t<0$.
Equation (4.6c) follows from eq.\ (4.6b) if we note
$\delta(t)\Delta({\bf x})=0$.  One can go directly from
eq.\ (4.6a) to eq.\ (4.6e) by doing the contour integral, however the
use of $i$ has been excluded.  The generator of the residue can possibly be
taken to be the geometry associated with $dk^4$, again the (minus)
time basis vector $-\basis{4}$ (see for example
Hestenes\cite{Hestenes68}) where there must be attention to its
placement.  Equivalently we state (without proof) that the unit
four-velocity $\phat$ may be used [inspection of eq.\ (4.6e) shows this
to be consistent].

Inspection of eq.\ (4.6e) shows that $\overline{G}({\bf x})=
G(-{\bf x})$, hence one can show that this Green function also
satisfies the adjoint equation:
$G({\bf x})(\Box - m) =\delta^4({\bf x})$, where $\Box$ is
now understood to operate to the left.  Hence we may derive the
general integral solution,
$$\Psi({\bf x}^{\prime})=
\oint d\Sigma^{\mu}\ G({\bf x^{\prime},x})
\ \basis{\mu}\ \Psi({\bf x}), \eqno(4.7)$$
where ${\bf x}$ is an interior point.  Actually, its abit more
complicated than that, one must add a homogeneous term to
eq.\ (4.6a) such that the boundry conditions are met.

\subsection{Path Integral Formulation}
The Feynman {\tt PIF} (Path Integral Formulation) of non-relativistic
quantum mechanics\cite{Feynman} provides an elegant means of
bridging the interpretation gap between classical mechanics and
quantum formulations. In the non-relativistic case, it asserts
that the propagator kernel of eq.\ (4.3a) can be written,
$$F({\bf x,x^{\prime}})=
\sum{\exp\left(\frac{i}{\hbar} {\cal S}
({\bf x,x^{\prime}})\right)}, \eqno(4.8a)$$
$${\cal S}({\bf x,x}^{\prime})=\int_{\bf x^{\prime}}^{\bf x}
p^{\mu}dx_{\mu}. \eqno(4.8b)$$
The {\it classical action} ${\cal S}$ is evaluated over the path
from ${\bf x^{\prime}}$ to ${\bf x}$.
The sum in eq.\ (4.8a) is over all possible classical paths:
$x^{\mu}=x^{\mu}(\tau)$, where $p^{\mu}=m \frac{dx^{\mu}}{d\tau}$
and $\tau$ is an affine parameter called the {\it proper time}.
Classical paths require $\frac{dx^4}{d\tau}>0$, i.e. paths in
which particles go backward in time are excluded.

Because of the second order time derivative, the Klein-Gordon
eq.\ (3.1) does not easily lend itself to {\tt PIF}.  The first
order Dirac eq.\ (3.2a) is a better candidate.  However, the
standard derivation of eq.\ (4.8a) from the Schr\"odinger wave
equation\cite{Abers} required that each eigenfunction have the same
generator of quantum phase, a commuting $i$ which is now unavailable
to us in the real multivector theory.  In fact, we have seen from
eq.\ (3.5a) that each eigenfunction (as seen from the left side) has
its own unique quantum phase generator $\phat$, which makes
the {\tt PIF} derivation problematic.  In the restricted case
of plus parity eigenfunctions however, we saw from eq.\ (3.7a)
that they all have the \underline{same} generator $\basis{4}$ when
viewed from the \underline{right} side.  We exploit the advantage
of multivector wavefunctions (over standard Dirac column spinors)
in that we can now consider \underline{right-side applied} operations.
Hence we introduce here a new alternative to eq.\ (4.3a), the
{\it dextrad propagator},
$$\Psi(\tpo{X}{t})= \int d^3\vector{X}^{\prime}
\ \Psi(\tpo{X^{\prime}}{t^{\prime}})
\ H(\tpo{X}{t};\tpo{X^{\prime}}{t^{\prime}}), \eqno(4.9a)$$
$$H({\bf x,x^{\prime}})=H({\bf x-x^{\prime}})=
\int{ d^3\vector{P} \over {(2\pi)^3}}
\ \exp{[-\basis{4}\ p^{\mu}(x_{\mu}-x^{\prime}_{\mu})}]. \eqno(4.9b)$$
It can be easily shown\cite{Pezz8301} that this propagator may
be written in the {\tt PIF} form of eq.\ (4.8a) with the
replacement: $i\rightarrow -\basis{4}$.  Further, the other
half of the solution space, the negative parity solutions
(`antiparticles') will follow the same relation, except requiring
the replacement $\basis{4}\rightarrow\ -\basis{4}$ in eq.\ (4.9b)
and $i \rightarrow +\basis{4}$ in eq.\ (4.8a).

\section{Summary}
We are not aware of any other work in Clifford analysis that has
considered the {\tt PIF} (Path Integral Formulation) of kernels.
This approach suggests that the particular form of eq.\ (3.7a) is
the ``special'' multigeometric entity which propagates (unchanged)
in the manner to which we ascribe the term ``quantum particle''.
Further, the {\tt PIF} provides a method to introduce
interactions into the quantum equation based upon classical mechanics.
Because of eq.\ (4.9a), we see that the interactions will couple
{\it dextrally} to the wavefunction (multiplied on the right).
In particular, the generator of electromagnetic interactions will be
$\basis{4}$ applied on the right\cite{Pezz9302}.
Finally, separate from these physical interpretations, the general
mathematical relationship between the  ``right-side applied''
{\it dextrad kernel} introduced in
eq.\ (4.9a) and the standard ``left-side applied'' one of eq.\ (4.3a)
[both as an integral solutions to eq.\ (3.2a)] should be explored.
In particular, it is unknown to the author if a general `dextrad'
form of eq.\ (4.7) exists, nor under what conditions it would
reduce to eq.\ (4.9a).

\bigskip
{\noindent \large  \bf Acknowledgements}\\
The author wishes to thank the organizers of the conference for the
arrangement of travel support for himself, and two graduate students:
John Adams (Physics, San Francisco State U.) and Matthew Enjalran
(Physics, Univ. of Mass., Amherst).  Numerous conversations
with J. Ryan G. Sobczyk and M. Shapiro have helped the author
in matching unfamiliar mathematical terminology with physical concepts.
Thanks to to Craig Harrison (Philosophy Dept., San Francisco State U.)
for discussions on the interface of mathematics and physics,
and in particular for pointing out reference\cite{Wigner}.
Finally, thanks to thesis advisor G. Erickson (Physics, Univ. of
Calif., Davis), who some 10 years ago guided the author in the
development of the original ideas on which this paper is based.

\end{document}